# Atomically thin MoS$_2$: A new direct-gap semiconductor

Kin Fai Mak[1], Changgu Lee[2], James Hone[2], Jie Shan[3], and Tony F. Heinz[1*]

[1] *Departments of Physics and Electrical Engineering, Columbia University, 538 West 120$^{th}$ St., New York, NY 10027, USA*

[2] *Department of Mechanical Engineering, Columbia University, New York, NY 10027*

[3] *Department of Physics, Case Western Reserve University, 10900 Euclid Avenue, Cleveland, OH 44106, USA*

Abstract

The electronic properties of ultrathin crystals of molybdenum disulfide consisting of $N = $ 1, 2, ... 6 S-Mo-S monolayers have been investigated by optical spectroscopy. Through characterization by absorption, photoluminescence, and photoconductivity spectroscopy, we trace the effect of quantum confinement on the material's electronic structure. With decreasing thickness, the indirect band gap, which lies below the direct gap in the bulk material, shifts upwards in energy by more than 0.6 eV. This leads to a crossover to a direct-gap material in the limit of the single monolayer. Unlike the bulk material, the MoS$_2$ monolayer emits light strongly. The freestanding monolayer exhibits an increase in luminescence quantum efficiency by more than a factor of 1000 compared with the bulk material.

The transition-metal dichalcogenide semiconductor $MoS_2$ has attracted great interest because of its distinctive electronic, optical, and catalytic properties, as well as its importance for dry lubrication [1-19]. The bulk $MoS_2$ crystal, an indirect-gap semiconductor with a band gap of 1.29 eV [20], is built up of van-der-Waals bonded S-Mo-S units [2, 5, 6, 11]. Each of these stable units consists of two hexagonal planes of S atoms and an intermediate hexagonal plane of Mo atoms coordinated through covalent interactions with the S atoms in a trigonal prismatic arrangement (Fig. 1a). Because of the relatively weak interactions between these layers and the strong interlayer interactions, the formation of ultrathin crystals of $MoS_2$ by the micromechanical cleavage technique is possible, as demonstrated by Novoselov et al. [21].

In this Letter, we report a systematic study of the evolution of the optical properties and electronic structure of ultrathin crystals of $MoS_2$ as a function of thickness for $N$ = 1, 2, 3, ... 6 layers of the material. Our study has investigated samples prepared both on solid surfaces and as free-standing films, entirely unperturbed by substrate interactions. The properties of the films were examined using three complementary spectroscopic techniques: optical absorption, photoluminescence (PL), and photoconductivity, with additional sample characterization provided by atomic-force microscopy (AFM). The combination of these spectroscopic methods allowed us to trace the evolution of both the indirect and direct band gaps of the material as a function of the layer thickness $N$. With decreasing thickness, the experiments reveal a progressive confinement-induced shift in the indirect gap from the bulk value of 1.29 eV up to 1.90 eV. The change in the indirect-gap energy was found to be significantly larger than that of the direct gap, which increased by only 0.1 eV. As a consequence of these different scaling properties and as suggested by previous calculations [8, 9], the MoS$_2$ crystals were seen to undergo a crossover from an indirect gap semiconductor to a direct gap material in the monolayer limit. In addition to the signatures of this effect in the absorption and photoconductivity spectra, the luminescence quantum yield (QY) showed a dramatic enhancement in going from the dark, indirect-gap bulk crystal to the bright, direct-gap monolayer. For our suspended samples, we observed an increase of the PL QY by more than a factor of 1000 for the monolayer compared with the bulk crystal.

We prepared mono- and few-layer MoS$_2$ samples from bulk MoS$_2$ (SPI Supplies) using a mechanical exfoliation technique similar to that employed for graphene [21]. Typical samples ranged from 25 to 200 μm$^2$ in size. Suspended samples were obtained by depositing the MoS$_2$ layers on oxide-covered Si substrates that had been prepared with arrays of holes (1.5 and 0.75 μm in diameter) [22, 23]. The arrays were patterned by nanoimprint lithography followed by CF$_4$/O$_2$ plasma etching. We identified areas with thin MoS$_2$ samples with an optical microscope (Fig. 2a). Photoluminescence images were also collected to assess sample quality (Fig. 2b). The sample thickness was analyzed by atomic-force microscopy (AFM). The reflectance contrast, *i.e.,* the ratio of the reflected radiation from the sample on the substrate to that from the bare substrate, exhibits a linear increase in magnitude with number of monolayers $N \leq 6$ and can also be used to determine the sample thickness with monolayer accuracy. More details of sample preparation and thickness characterization are provided in Auxiliary Material 1 [24].

Optical measurements by absorption, PL, and photoconductivity spectroscopy were performed on mono- and few-layer MoS$_2$ samples. All optical measurements were performed under ambient conditions at room temperature using a Nikon inverted microscope coupled to a grating spectrometer with a CCD camera. The optical beams were focused on the sample with a spot diameter of ~ 1 μm. Details are given in Auxiliary Material 2 [24]. Briefly, for absorption measurements [25], samples on transparent fused quartz substrates were studied in the photon energy range of 1.3 – 3.0 eV using a quartz-tungsten-halogen source. For PL studies, suspended samples were excited with a cw solid-state laser at a wavelength of 532 nm. A low laser power of ~ 50 μW (on the sample) was used to avoid heating and PL saturation. The PL QY of $MoS_2$ samples was calibrated using a thin film of rhodamine 6G of about 50 nm thickness as a PL standard. The QY of the rhodamine 6G thin film was determined independently by comparing it to that of a dilute solution of rhodamine 6G (≤ 300 μM) in methanol. The fluorescence QY of the latter is known to be close to 1. For the photoconductivity studies, two-terminal devices were fabricated from mono- and bilayer $MoS_2$ samples on oxide-covered Si substrates by means of e-beam lithography. Laser radiation of well-defined frequency was selected from a super-continuum source with the use of a monochromator. The radiation was modulated with a mechanical chopper and the synchronous signal was detected with a lock-in amplifier. The spectral dependence of the photoconductivity was obtained from the photocurrent normalized by the intensity of the optical excitation reflected from the substrate.

As an indirect gap material, band-gap photoluminescence (PL) in bulk $MoS_2$ is a weak phonon-assisted process and is known to have negligible QY. Appreciable PL was, however, observed from few-layer $MoS_2$ samples, and surprisingly bright PL was detected from monolayer samples. The PL intensity measured at room temperature under identical excitation at 2.33 eV for a suspended monolayer and a bilayer sample is strikingly different (Fig. 3a). The luminescence QY for suspended samples of $N = 1 - 6$ thickness drops steadily with increasing thickness (Fig 3a, inset). A PL QY on the order of $10^{-5} - 10^{-6}$ was estimated for few-layer samples of $N = 2 - 6$; a value as high as $4 \times 10^{-3}$ was observed when the sample reached the limit of monolayer thickness. In addition to the significant difference in the luminescence QY, the PL spectra for mono- and few-layer samples are quite distinct from one another (Fig. 3b). The PL spectrum of suspended monolayer samples consists of a single narrow feature of ~ 50 meV width that is centered at an energy of 1.90 eV. In contrast, few-layer samples display multiple emission peaks (A, B, and I). Peak A coincides with the monolayer emission peak. It redshifts and broadens slightly with increasing $N$. Peak B lies about 150 meV above peak A. The broad feature I, which lies below peak A, shifts systematically to lower energies and becomes less prominent with increasing $N$. Its position for $N = 2 - 6$ layers approaches the indirect gap energy of 1.29 eV (Fig. 3c).

To understand the origin of these extraordinary PL properties, we compare the PL to the absorption spectrum of $MoS_2$. In the photon energy range of 1.3 - 2.4 eV, the absorption of bulk $MoS_2$ at room temperature is dominated by two prominent resonance features. They are known to arise from direct-gap optical transitions between the maxima of split valence bands ($v1$, $v2$) and the minimum of the conduction band ($c1$), all located at the K-point of the Brillouin zone (Fig. 1b) [7, 12-18]. The splitting arises from the combined effect of the interlayer coupling and spin-orbit coupling. (The latter is not included in the band structure of Fig. 1b.) The absorption spectrum of atomically thin layers of $MoS_2$ was found to be largely unaltered, except for a slight blue-shift of the resonances (Fig. 4a). The monolayer PL peak A at 1.90 eV is seen to match the lower absorption resonance in both its position and width. Therefore, we attribute the PL of monolayer $MoS_2$ to direct-gap luminescence. The result, combined with the high PL QY, also demonstrates the excellent quality of the suspended monolayer samples. For bilayer samples, on the other hand, emission peaks A and B match the two absorption resonances and are assigned to direct-gap hot luminescence (for details, refer to Auxiliary Material 3 [24]). The weak feature between A and B could be impurity or defect luminescence and merits further investigation. The PL peak I at 1.59 eV (~ 300 meV below the direct-gap absorption peak) shifts systematically with the layer number (Fig. 3c). As we discuss below, it is attributed to indirect-gap luminescence.

Direct probing of the indirect-gap optical transitions in atomically thin layers of $MoS_2$ is limited by the sensitivity of absorption measurements. In the case of weak exciton binding, as is for $MoS_2$ at room temperature, the photoconductivity spectrum mimics the absorption spectrum. Therefore, we used photoconductivity spectroscopy to investigate the optical response below the direct gap of mono- and bilayer $MoS_2$ [12]. The photoconductivity spectra of these samples could be recorded with signals ranging over three decades in strength (Fig. 4b). For bilayer $MoS_2$, the onset of photoconductivity occurs around 1.60 eV, coinciding with the PL peak I. The conductivity increases slowly with photon energy towards the direct band gap, as is characteristic of indirect-gap materials. In contrast, for monolayer $MoS_2$, well below the direct-gap transition no photoconductive response is seen; the photoconductivity exhibits an abrupt increase only near the direct band gap. These features suggest that while bilayer $MoS_2$ is an indirect-gap semiconductor, monolayer $MoS_2$ is a direct-gap material [26].

Such a conclusion is supported by analysis of the spectral dependence of the photoconductivity. Here we present a simplified treatment that aims only to capture the dominant factors. Accordingly, we neglect both excitonic effects and the variation of matrix elements with energy, factors that should be included in a more comprehensive theory. For a two-dimensional (2D) material such as thin layers of $MoS_2$, the absorbance $A(\hbar\omega)$ at photon energy $\hbar\omega$ for a direct band-gap transition of energy $E_g$ is determined by the joint density of states of the relevant bands and assumes a step-function-like dependence [1, 27] $A(\hbar\omega) \propto \Theta(\hbar\omega - E_g)$. Transition between indirect band valleys of gap energy $E'_g$, on the other hand, is a phonon-assisted process; the corresponding absorbance at room temperature can be estimated as [1, 27]

$$A(\hbar\omega) \propto \sum_\alpha \left[ \frac{\hbar\omega - \theta_\alpha - E'_g}{1 - \exp(-\theta_\alpha / kT)} + \frac{\hbar\omega + \theta_\alpha - E'_g}{\exp(\theta_\alpha / kT) - 1} \right] \propto \hbar\omega - E'_g$$

, where $\theta_\alpha$ denotes the energy of the $\alpha$-th phonon mode and $kT$ is the thermal energy. These simple dependences, with a broadening of 30 meV for the direct-gap transitions, were found to describe the experimental photoconductivity spectra quite well (Fig. 4b). For bilayer $MoS_2$, both indirect and direct-gap absorption are required to explain the onset of photoconductivity below the direct-gap energy. The direct-gap absorption alone is sufficient to account for the photoconductivity spectra in monolayer $MoS_2$.

Such an indirect-direct gap crossover, as mentioned above, is the result of a significant up-shift of the indirect gap energy induced by a strong quantum confinement effect. Let us examine the electronic band structure of bulk $MoS_2$ (Fig. 1a). The direct gap of ~ 1.8 eV occurs between the lowest conduction band ($c1$) and the highest valence split bands ($v1$ and $v2$) at the K-point of the Brillouin zone (transitions A and B) [3-5, 7, 12]. On the other hand, the valence band (*v1*) maximum and conduction band (*c1*) minimum are located at the Γ-point and along the Γ – K direction of the Brillouin zone [3-6, 8, 11], respectively; they form an indirect gap of 1.29 eV (transition I) [20]. The out-of-plane mass $m_\perp$ for electrons and holes around the K-point far exceeds the free electron mass $m_0$; $m_\perp$ for holes around the Γ-point is estimated to be ~ $0.4m_0$ and for electrons around the *c1* minimum along the Γ – K direction to be ~ $0.6m_0$. These significant differences in the out-of-plane particle masses, determined by dispersion of the bands in the out-of-plane direction, reflect the distinct nature of their electronic states [5]. Thus, while decreasing the layer number *N* leads to a significant confinement-induced increase in the indirect-gap energy, the direct band gap remains almost unchanged. The indirect-direct gap crossover occurs in the limit of monolayer thickness.

To explain this result more precisely, we apply a zone-folding scheme to produce the electronic states of the ultrathin samples [28, 29]. For identical layers with only nearest-neighbor interactions, the 2D electronic structure of few-layer MoS$_2$ samples can be generated as a subset of the bulk electronic states with quantized in-plane momenta, *i.e.,* momenta lying in planes perpendicular to the Γ-A or K-H directions in the Brillouin zone (for details, refer to Auxiliary Material 4 [24]). For MoS$_2$ of monolayer thickness, the allowed plane passes directly through the H and A points. For the resulting 2D bands, the conduction band minimum and the valence band maximum both occur at the H point (Fig. 1b). Monolayer MoS$_2$ therefore becomes a direct-gap material. With increasing layer number, cuts with out-of-plane momenta approaching the Γ – K line are made. Then the conduction band minimum and the valence band maximum occur at the Γ point and along the Γ – K direction, as in the bulk material. Few-layer MoS$_2$ samples are, therefore, indirect-gap semiconductors, like the bulk material. These findings agree with earlier density-functional theory (DFT) calculations of few-layer MoS$_2$ [8, 9].

The crossover from an indirect-gap material to a direct-gap material naturally accounts for the more than 1000-fold enhancement of the luminescence QY observed in monolayer MoS$_2$ (Fig. 3a). It is consistent with the observed absorption spectra that show little sensitivity to the layer number $N$ (Fig. 4a) and a band gap that decreases with increasing $N$ (Fig. 3c). The observed dependence of the band gap on $N$ is also in qualitative agreement with published DFT calculations [8]. The MoS$_2$ monolayer constitutes the first atomically thin material that is an effective emitter of light. The strong light emission from monolayer MoS$_2$ suggests its use for photostable markers and sensors that can be adapted to probe nanoscale dimensions. The controllability of the band gap may also be used to optimize the material's use as a photocatalytic agent [19] and for photovoltaic applications. Our results show that the distinctive electronic properties of ultrathin layered materials are not restricted to graphene [30], but extend to a broader group of van-der-Waals bonded solids. This will allow for a corresponding expansion of impact and applications of such atomically thin materials.

The authors acknowledge support from the National Science Foundation under grant CHE-0117752 at Columbia and grant DMR-0907477 at Case Western Reserve University; from DARPA under the CERA program; and from the New York State Office of Science, Technology, and Academic Research (NYSTAR).

**Figures**

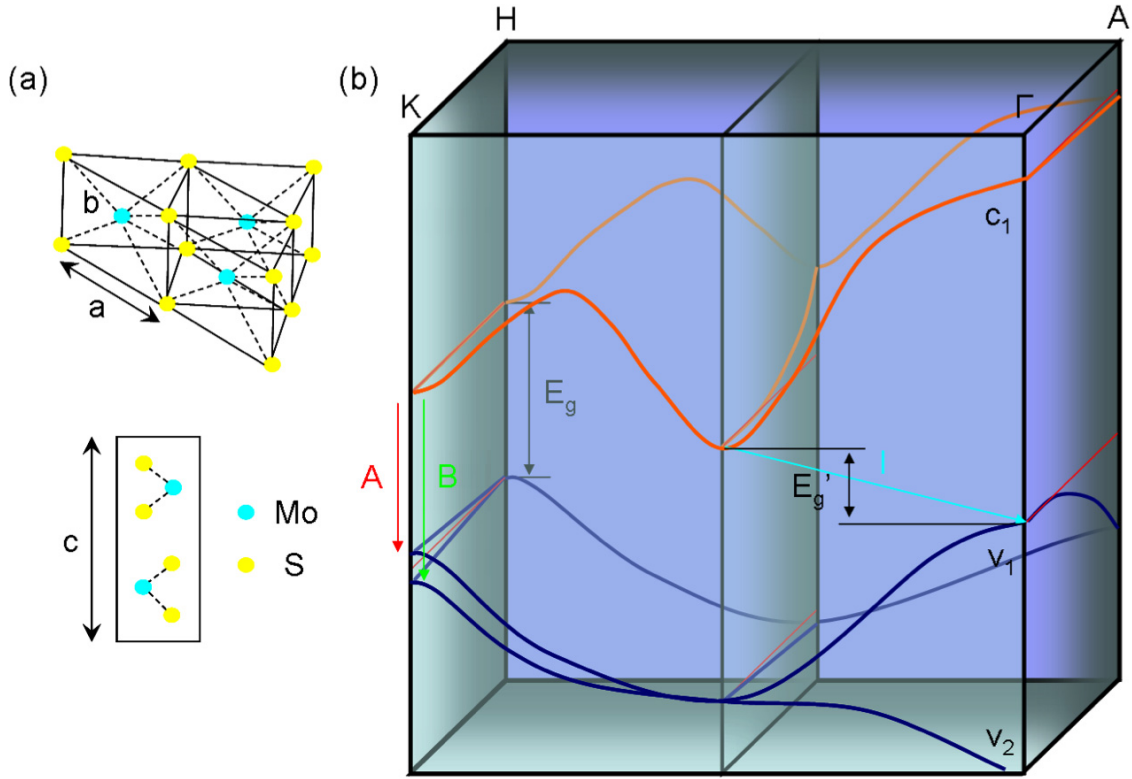

Figure 1. (a) Lattice structure of $MoS_2$ in both the in-plane and out-of-plane directions. (b) Simplified band structure of bulk $MoS_2$, showing the lowest conduction band $c1$ and the highest split valence bands $v1$ and $v2$. A and B are the direct-gap transitions and I is the indirect-gap transition. Also indicated are indirect gap for the bulk material $E_g'$ and the direct gap $E_g$ for the monolayer, the latter as predicted by the zone-folding analysis through isolating the AH plane.

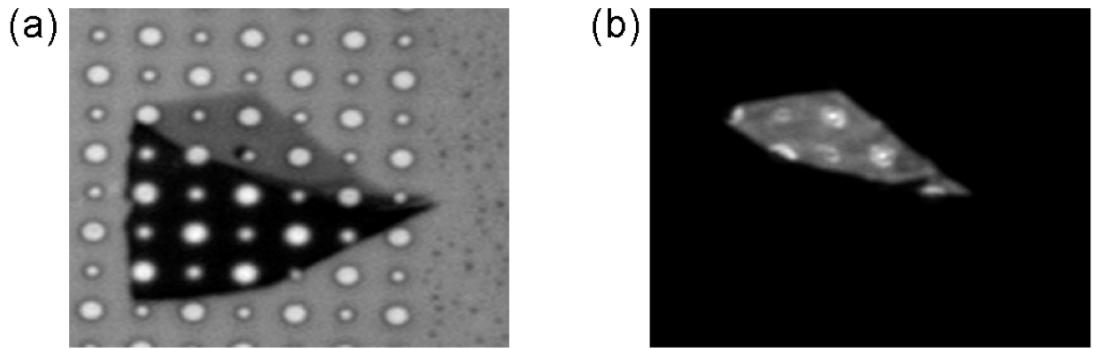

Figure 2. (a) Representative optical image of mono- and few-layer MoS$_2$ crystals on silicon substrate with etched holes of 1.5 and 0.75 μm in diameter. (b) PL image of the same samples shown in (a). The PL QY is much enhanced for suspended regions of the monolayer samples, demonstrating its high quality. Note that the emission from the few-layer sample is too weak to be seen in this contrast setting.

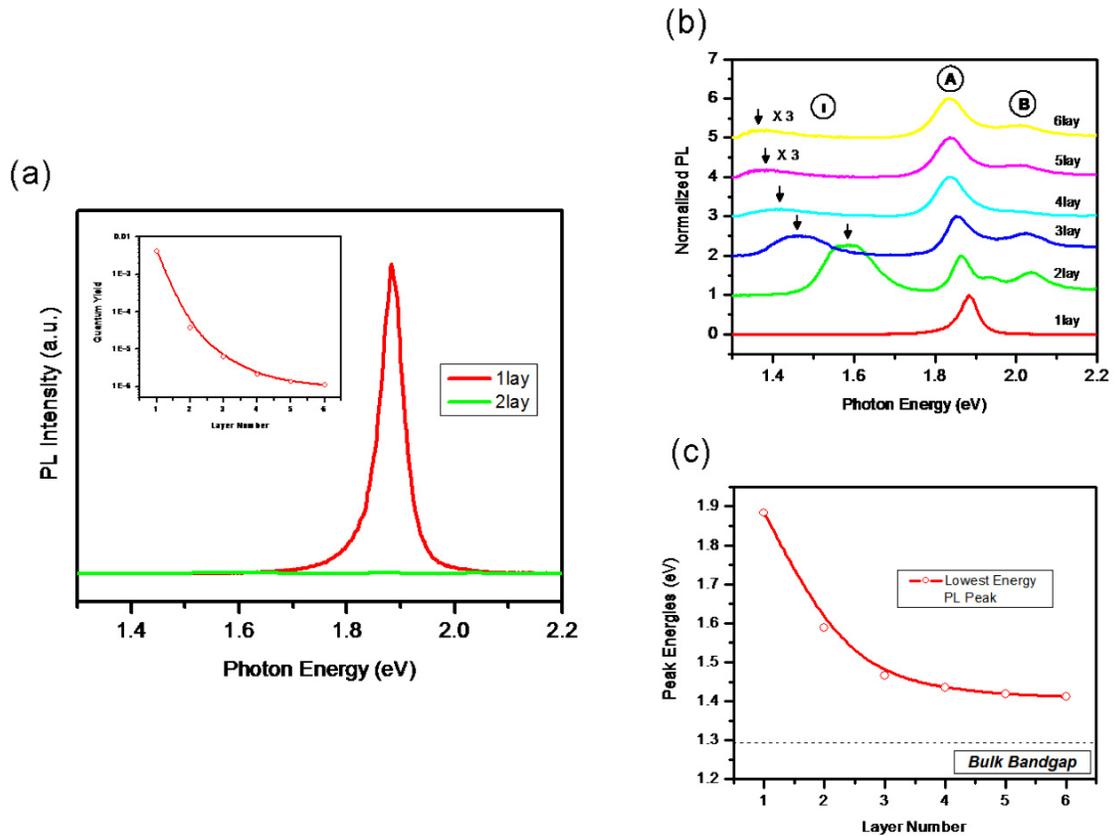

Figure 3. (a) PL spectra for mono- and bilayer MoS$_2$ samples in the photon energy range from 1.3 to 2.2 eV. Inset: PL QY of thin layers of MoS$_2$ for number of layers $N = 1 - 6$. (b) PL spectra of thin layers of MoS$_2$ for $N = 1 - 6$. The spectra are normalized by the intensity of peak A and are displaced for clarity. (c) Band-gap energy of thin layers of MoS$_2$ for $N = 1 - 6$. The band-gap values were inferred from the energy of the PL feature I for $N = 2 - 6$ and from the energy of the PL peak A for $N = 1$. As a reference, the (indirect) band-gap energy of bulk MoS$_2$ is shown as dashed line.

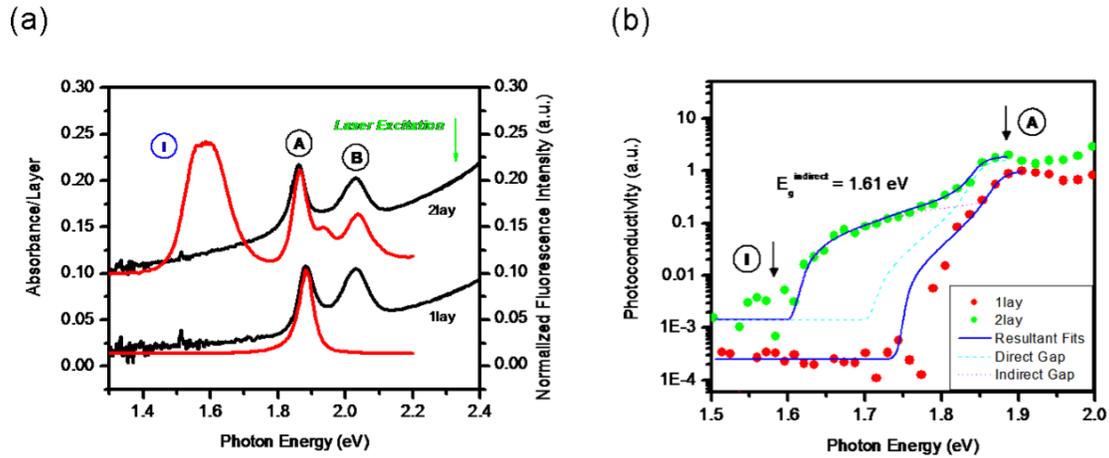

Figure 4. (a) Absorption spectra (left axis) normalized by the layer number *N* in the photon energy range from 1.3 to 2.4 eV. The corresponding PL spectra (right axis, normalized by the intensity of the peak A) are included for comparison. The spectra are displaced along the vertical axis for clarity. (b) Photoconductivity spectra for mono- (red dots) and bilayer (green dots) samples. The data are compared with the spectral dependence predicted for 2D semiconductors using the model described in the text (blue lines). Contributions from both indirect-gap (pink dotted line) and direct-gap (green dotted line) transitions are required to describe the photoconductivity spectrum of the bilayer; a direct-gap transition alone is adequate to describe the photoconductivity spectrum of the monolayer. For comparison, the energies of the direct and indirect gap inferred from the PL measurements are indicated by arrows.